\documentclass[aps,prl,twocolumn,showpacs,superscriptaddress,amsmath,amssymb]{revtex4-1}
\usepackage{graphicx}
\usepackage{color,soul}
\usepackage{xcolor}

\newcommand{\nvec}[0]{{\bm n}}
\newcommand{\rvec}[0]{{\bm r}}
\newcommand{\Evec}[0]{{\bm E}}

\begin{document}

\title{Convolutional Neural Network analysis of optical texture patterns in liquid-crystal skyrmions}

\author{J. Terroa} 
\affiliation{Centro de F\'{i}sica Te\'{o}rica e Computacional, Faculdade de Ci\^{e}ncias, Universidade de Lisboa, 1749-016
Lisboa, Portugal} 

\author{M. Tasinkevych}
\affiliation{Department of Physics and Mathematics, School of Science and Technology, Nottingham Trent University, Clifton Lane, Nottingham NG11~8NS, United Kingdom.} 
\affiliation{International Institute for Sustainability with Knotted Chiral Meta Matter (WPI-SKCM$^2$), Hiroshima University, Higashi-Hiroshima, Hiroshima 739-8526, Japan}

\author{C. S. Dias} \email{csdias@fc.ul.pt}
\affiliation{Centro de F\'{i}sica Te\'{o}rica e Computacional, Faculdade de Ci\^{e}ncias, Universidade de Lisboa, 1749-016
Lisboa, Portugal} 
\affiliation{Departamento de F\'{\i}sica, Faculdade de Ci\^{e}ncias,
Universidade de Lisboa, 1749-016 Lisboa, Portugal}

\begin{abstract}
Liquid crystals are known for their optical birefringence, a property that gives rise to intricate patterns and colors when viewed in a microscope between crossed polarisers. Resulting images  are rich in geometric patterns and serve as valuable fingerprints of the liquid crystal's intrinsic properties. By using machine learning techniques, it is possible to extract from the images information about, e.g., liquid crystal elastic constants, the scalar order parameter, local orientation of the director, etc. Machine learning can also be employed to identify phase transitions and classify different liquid crystalline phases and topological defects. In addition to well studied singular defects such as point or line disclinations, liquid crystals can also host non-singular solitonic defects such as skyrmions, hopfions, and torons. The solitons, with their localised and stable configurations, offer an alternative view into material properties and behaviour of liquid crystals. In this study, we demonstrate that the optical signatures of skyrmions can be utilised effectively in machine learning to predict important system parameters. Our method focuses specifically on the skyrmion-localised regions, reducing significantly the computational cost. By training convolutional neural networks on simulated polarised optical microscopy images of liquid crystal skyrmions, we showcase the ability of trained networks to accurately predict several selected parameters such as the free energy, cholesteric pitch, and strength of applied electric fields. This study highlights the importance of localized topologically arrested order parameter configurations for materials characterisation research empowered by state-of-the-art data science methods, and may pave the way for the development of advanced skyrmion-based applications. 

\end{abstract}

\maketitle

\section*{Introduction}

Liquid crystals (LCs) are a unique state of matter, exhibiting fluid-like behaviour while maintaining the orientational order of their constituent anisotropic molecules \cite{Collings2017, deGennes1993}. This fascinating duality gives rise to remarkable optical properties, making LCs the cornerstone of the multi-billion-dollar display industry. The underlying orientational order, easily influenced by boundaries or external stimuli, results in intricate intensity patterns observed between crossed polarisers in polarised optical microscopy (POM) images \cite{Lagerwall2006,Oswald2005}. Within these complex patterns, an important role is played by topological defects, singular points or lines where the orientational order is disrupted, offering a unique window into the fundamental properties of LCs\cite{Demus1998}. Topological defects, with their characteristic configurations and dynamics, can serve as sensitive probes of material parameters and external influences. In this work, we aim to leverage the rich information encoded in the textures surrounding a specific type of non-singular topological defects$-$the skyrmions$-$to predict various LC properties using convolutional neural networks (CNNs).

Skyrmions, in the context of LCs, are particle-like, two-dimensional topological defects within the alignment field of rod-like molecules \cite{Ackerman2017}. The corresponding alignment (director) field possesses a distinctive topologically protected structure characterised by an integer-valued skyrmion number, signifying the number of times the director field wraps around the order parameter manifold \cite{Sohn2019}. The topological protection ensures skyrmion stability against external stimuli, such as moderate electric fields or temperature variations \cite{Porenta2014a}. Skyrmions exhibit fascinating dynamics, including electrically-driven motion and the ability to form dynamic clusters, resembling the schooling behaviour observed in active matter \cite{Foster2019}. These properties make them relevant for potential applications in microfluidics, racetrack memory devices, and other emerging technologies \cite{Porenta2014a}.

A promising unexplored avenue for predicting material properties is offered by analysing the POM textures due to the skyrmions. By observing how skyrmions respond to electric fields or other perturbations, one can indirectly measure material parameters such as elastic constants, dielectric anisotropy, viscosity \cite{Smalyukh2020}, etc. This approach could be particularly valuable for the LC display industry, where precise control and understanding of material behaviour are crucial for optimising the device performance and developing new display technologies \cite{Sohn2020a}. The skyrmions are not only stable due to their topological protection, but also highly localized in space \cite{Porenta2014a}. Despite their relatively small size$-$of the order of several micrometers$-$the skyrmion director configurations contain a wealth of information \cite{Smalyukh2020}, which may be decoded form the skyrmion POM images by using pre-trained CNNs.

Machine learning methods have been  employed extensively to predict various LC characteristics, including elastic constants \cite{Zaplotnik2023} and dielectric anisotropy \cite{Taser2023} as well as to identify LC phases \cite{Piven2024}. Among these techniques, CNNs have been proven effective in learning properties of LCs from their optical textures \cite{Pessa2022,Sigaki2020,Smith2020}. CNNs have also been used to investigate the dynamics of optical patterns of LC films exposed to gas mixtures in order to detect the presence of specific species \cite{Bao2022}. Beyond leveraging POM textures, machine learning can also utilise molecular structure of mesogens to predic LC properties \cite{Chen2019,Inokuchi2020}. The LC molecular orientation field can host a large variety of topological point and line defects where the director field is discontinuous. Defects indentification, classification and analysis provide additional insights into LC material behaviour \cite{Walters2019,Minor2020,Sakanoue2021}.

From this perspective, CNNs offer the advantage of being able to capture spatial correlations and identify intricate patterns associated with defects, which enables a more detailed analysis. Indeed, CNNs have been used not only to classify various LC phases but also to distinguish between subtle textural differences, demonstrating potential in designing automated material characterisation applications \cite{Dierking2022,Dierking2023,Dierking2023a}.

In this work, we exploit the capability of CNNs to infer LC material properties from  skyrmion optical signatures. By training the networks on POM textures produced due to the skyrmions, we aim to extract not only intrinsic LC properties such as the cholesteric pitch but also information about external influences, e.g., electric fields. This approach may facilitate characterisation of LCs and pave the way for advancements in display technologies. Notably, despite their localised nature LC skyrmions encapsulate rich information within their director configurations which may be pooled out from POM images, making the skyrmions valuable probes for materials research.

\section*{Results}

LC skyrmions have been realised experimentally in chiral nematic liquid crystals confined between two parallel plates, imposing perpendicular boundary conditions on the LC director field $\nvec(\rvec)$\cite{Fukuda2011}. Chiral or cholesteric LCs are materials displaying a periodic helical (chiral) structure of $\nvec(\rvec)$. It can be visualised as a stack of nematic layers, each having a well-defined director orientation, which changes/twists periodically across the layers. The axis which is perpendicular to the layers is called the helical axis, and the distance along the axis over which the director twists by 360° is known as the cholesteric pitch, $\eta$\cite{Sigaki2020}. The value of $\eta$ can be estimated easily when the helical axis is perpendicular to the viewing direction of an optical microscope but cannot be obtained from standard experimental arrangements used in reflective displays\cite{Zheng2017a}. Therefore, it is of practical interest to find a simple way of extracting  $\eta$ directly from skyrmion samples. To tackle this issue, we will use CNNs trained on numerically computed POM textures.

\subsection*{Single-skyrmion data generation}

To effectively train any type of Machine Learning model, a sufficient amount of training data is crucial. In this study, we generate the necessary training data by minimising the Frank-Oseen elastic free energy denoted as $F$ to accurately describe the behaviour of LC under the influence of external electric fields. The system geometry and details of the numerical implementation are described in the Methods section. Under strong perpendicular boundary conditions, we obtain director profiles with so-called torons \cite{Smalyukh2010}, which may be visualised as skyrmion tubes which terminate in the vicinity of the confining surfaces on point defects. This allows embedding of the skyrmion tube in a uniform background director field in three dimensions (3D). 
For the skyrmion tube, two-dimensional (2D) director profiles within each of the cross-sections perpendicular to the tube axis has the structure of the baby skyrmion where the director twists uniformly by 180$^\circ$ from the tube center in all radial directions \cite{Ackerman2017}. 

The minimisation process renders 3D director configurations, which are then mapped to the POM images required for training of the CNNs (see Methods section for a detailed explanation). We generate single wavelength POM textures from the free energy minimising $\nvec(\rvec)$ by applying the standard Jones matrix method \cite{Zhao2023,Yeh2009}. The resulting optical image sums up, in an effective way, the configurational information stemming from all the skyrmions which exist within the 2D LC layers perpendicular to the direction of light propagation.    

\begin{figure*}[thb]
    \centering
    \includegraphics[width=0.7\textwidth]{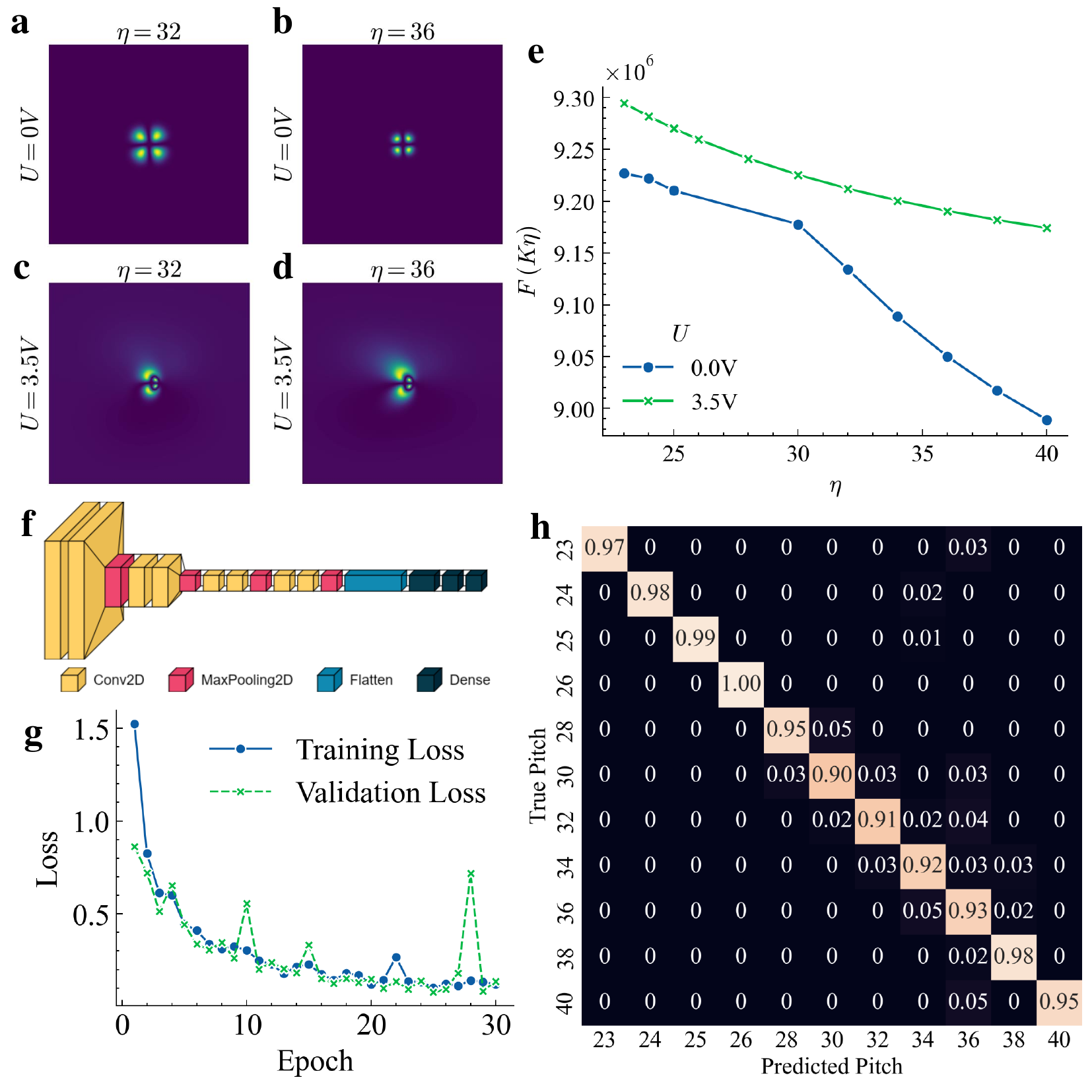}
    \caption{\textbf{a-d}. POM images corresponding to the director configurations obtained by numerical minimisation of the Frank-Oseen free energy at $\eta=32$ (\textbf {a, c}) and $\eta=36$ (\textbf {b, d}), under the presence of $U=0V$ (\textbf {a, b}) and  $U =3.5 V$ (\textbf {c, d}) electric fields. \textbf{e}. Frank-Oseen free energy as a function of $\eta$ at zero voltage (blue solid circles) and in electric field corresponding to $U=3.5V$ (green crosses). \textbf{f}. Schematic representation of the CNN architecture used for the classification of the POM images, where the yellow layers are the convolutional layers, the red ones the Max pooling layers, the light blue the flatten layers, and the dark blue the dense layers. \textbf{g}. Training and validation losses as functions of training epochs. \textbf{h}. Confusion matrix for the classification of $\eta$.}
    \label{fig:1}
\end{figure*}

Figure \ref{fig:1} illustrates the impact of both LC intrinsic properties and external perturbations on the skyrmion POM textures. More specifically, Figs.~\ref{fig:1}\textbf{a-d}  demonstrate how the texture changes by varying $\eta$ and applied voltage $U$. Figure \ref{fig:1}\textbf{e} depicts the free energy of the stabilised skyrmion structure as a function of $\eta$ for two values of $U$ (see Methods for details). The observed break in slope in the blue curve at $\eta = 30$ corresponds to the configurational transition where the uniform (without the torons) director profiles are stable for $\eta<30$, and non-uniform profiles with torons are observed at $\eta \ge 30$.

\subsection*{Single-skyrmion pitch prediction}

 To predict the pitch of the chiral nematic LC we have used the POM textures around  skyrmions. We have built a data set of skyrmions associated with different values of $\eta$ $\in$ \{23, 24, 25, ..., 40\} from the numerical minimisation of the Frank-Oseen free energy. Figure \ref{fig:1}\textbf{f} illustrates the network architecture used initially for this task. In this network, input images (300 × 300 pixels) pass through three blocks of two 4 × 4 convolutions and one 3 × 3 max-pooling layers, followed by three fully connected layers (with 32, 32, and 16 nodes, respectively) and an output layer. We have used rectified linear unit (ReLU) activation functions in all convolutional and fully connected layers, while the output layer uses a softmax activation function with 11 nodes. We have separated 15\% of the data for a final evaluation (test set) of the model and have used the remaining 85\% as a training set. The network parameters have been optimised using the Adam algorithm\cite{Kingma2014} (learning rate of 0.001), and the loss function is categorical cross-entropy (commonly used in multinomial classification). To avoid overfitting, we have applied an early stopping regularisation procedure (with patience set to 5 epochs) over all convolutional and fully connected layers. In Fig.~\ref{fig:1}\textbf{g}, we show how the loss function (given by the mean square error) evolves with the number of epochs. From comparing the training and the validation loss functions, we conclude that the CNN is not overfitted for this data. Figure \ref{fig:1}\textbf{h} shows the confusion matrix obtained by applying the trained network to the 15\% of data never exposed to the algorithm. These results demonstrate the excellent accuracy that our network can achieve in identifying the pitch of a skyrmion in the test set, where the average accuracy is $\approx0.96$.

\subsection*{Single-skyrmion electric field prediction} 

\begin{figure*}[thb]
    \centering
    \includegraphics[width=0.8\textwidth]{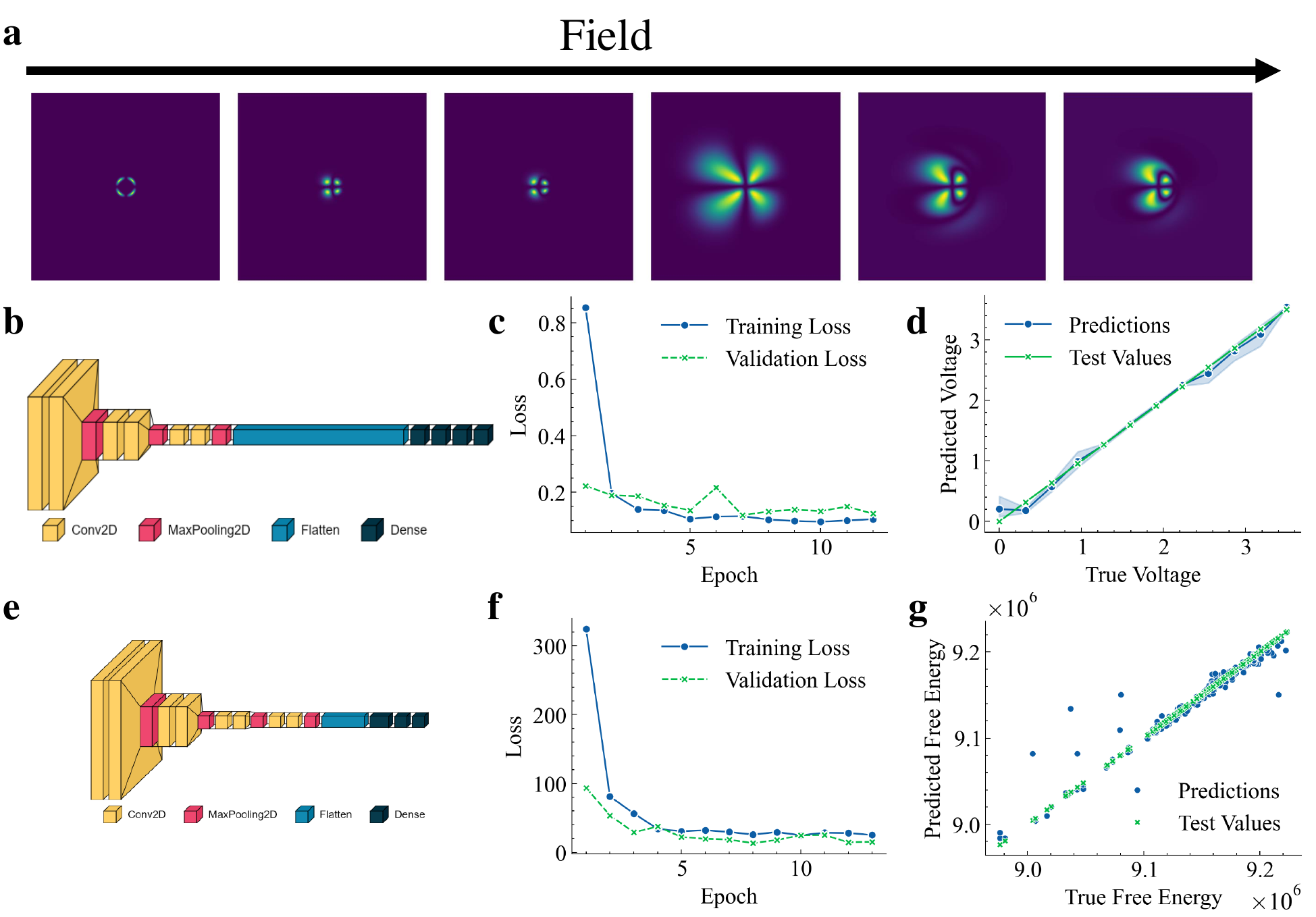}
    \caption{\textbf{a} POM images corresponding to the director configurations obtained by numerical minimisation of the Frank-Oseen free energy at $\eta=37$, for an applied electric field, from left to right of $U=0V$ to $U=3.5V$. \textbf{b} Network architecture\cite{Gavrikov2020} used for learning and estimating $U$ from the POM images. It is composed of three blocks of two convolutional and one max-pooling layers followed by four fully connected layers with 32 nodes each and an output layer with a single node and linear activation function. \textbf{c} Training and validation losses as functions of training epochs. \textbf{d} Relationship between the predicted and true voltage values obtained by applying the training network to the test set (the crosses is the 1:1 relationship). The blue shaded area represents the standard deviation for each point. \textbf{e} The network architecture\cite{Gavrikov2020} used for predicting the free energy of a skyrmion sample. Network with the similar general structure as in Fig.~ \ref{fig:1}f, now with an extra block of two convolutional (yellow) and one max-pooling (red) layers, a decrease to two fully connected layers with 32 and 16 nodes and the last layer is now composed of 1 node with a linear activation function. \textbf{f} Training and validation losses as functions of training epochs. \textbf{g} Relationship between predicted and true energy values obtained by applying the training network to the test set (the crosses is the 1:1 relationship). The blue shaded area represents the standard deviation for each point.}
    \label{fig:2}
\end{figure*}

We now demonstrate the feasibility of using POM textures of isolated skyrmions to predict the intensity of an applied electric field. Differently from the pitch classification which is an integer number, it is now a regression problem where the network output is a continuous variable representing the applied voltage $U$. Figure~\ref{fig:2}\textbf{a} presents a series of POM images of a single skyrmion at a fixed $\eta$, and for increasing $U$. A transformation in the texture surrounding the skyrmion is evident, with even the size of the skyrmion exhibiting a dependence on $U$. This observation highlights the sensitivity of the skyrmion textures to external stimuli, suggesting their potential as probes for characterising such influences.

 We have employed the CNN architecture depicted in Fig.~\ref{fig:2}\textbf{b} to predict $U$ from the POM textures similar to those in Fig.~\ref{fig:2}\textbf{a}. The network consists of convolutional layers to extract features, followed by fully connected layers for regression. We have increased the number of blocks of two 4 × 4 convolutions and one 3 × 3 max-pooling layers by one, reduced the number of fully connected layers by one and replaced the softmax activation function of the output layer by a linear activation function. A ReLU activation functions have been used after all convolution operations. We have trained this network by optimising the mean square error (loss function), following the same procedures  used for the pitch classification. Figure~\ref{fig:2}\textbf{c} shows the evolution of the mean square error as the loss function during the training and testing phases, indicating that the CNN effectively learns the relationship between the POM textures and the applied voltage without overfitting. The performance of the trained CNN is illustrated in Fig.~\ref{fig:2}\textbf{d}, where the predicted $U$ is plotted against the true values. The close agreement with the 1:1 line and a coefficient of determination of approximately 0.94 demonstrate the high accuracy of our method in recovering the true voltage. This result evidences the fact that the CNNs can utilise effectively the information encoded within LC solitonic defects such as skyrmions to evaluate external stimuli applied to the system.

\subsection*{Single-skyrmion free energy prediction}

Predicting the free energy of a liquid crystal system is of importance for understanding the stability and dynamics of topological defects such as skyrmions.  By accurately estimating the free energy, we can anticipate how these defects will respond to changes in experimental parameters such as temperature, applied voltage, or confinement. This capability is relevant not only for fundamental research but also for the development of future skyrmion-based devices. The ability to extract the free energy directly from POM textures using machine learning can aid the optimisation of device parameters and accelerate the design process.

We have employed the CNN architecture shown in Fig.~\ref{fig:2}\textbf{e}, which is similar to the one (Fig.~ \ref{fig:1}f) used for classifying the skyrmion pitch. This architecture enables the effective capture of intricate features within the POM images and their relation to the system's free energy landscape. The network comprises four blocks of two 4x4 convolutional layers with a 3x3 max-pooling layer, ReLU activation functions after all convolution operations, and a linear activation function in the output layer. We have used 80\% of the data for training, 15\% for validation, and 5\% for final testing. The dynamics of the training process is presented in Fig.~\ref{fig:2}\textbf{f}, which displays the loss functions dependence on the training epochs. Both the training and testing losses decrease steadily, indicating successful learning without overfitting.  In Fig.~\ref{fig:2}\textbf{g} we highlight the accuracy of this approach, with the predicted free energy values closely matching the true values with a coefficient of determination of $\approx0.96$.

\subsection*{Two-skyrmion data generation}

The ability to predict liquid crystal properties in systems with multiple skyrmions holds significant promise for advancing both fundamental research and technological applications. "Schools" of skyrmions, exhibiting collective behaviour and intricate interactions, have garnered considerable attention due to their potential in areas such as novel displays with dynamic light control and efficient drug delivery systems \cite{Sohn2019,Sohn2018,Sohn2019a}. It is essential to extend our machine learning approach to handle scenarios with multiple skyrmions to accurately characterise and exploit these complex systems.

In addition to its relevance for applications, predicting properties in multi-skyrmion systems is crucial for understanding the fundamental nature of emergent collective skyrmion dynamics governed by out-of-equilibrium interactions. The later appear due to the elastic director distortions and topological constraints, and play a key role in determining the self-assembling behaviour and stability of skyrmion ensembles. To investigate these interactions, we have generated a dataset of POM images featuring pairs of skyrmions at various relative positions and pitch values, and under different applied voltages. This dataset enables us to explore the interplay between skyrmion separation, material properties, and external stimuli.

\begin{figure*}[thb]
    \centering
    \includegraphics[width=0.9\textwidth]{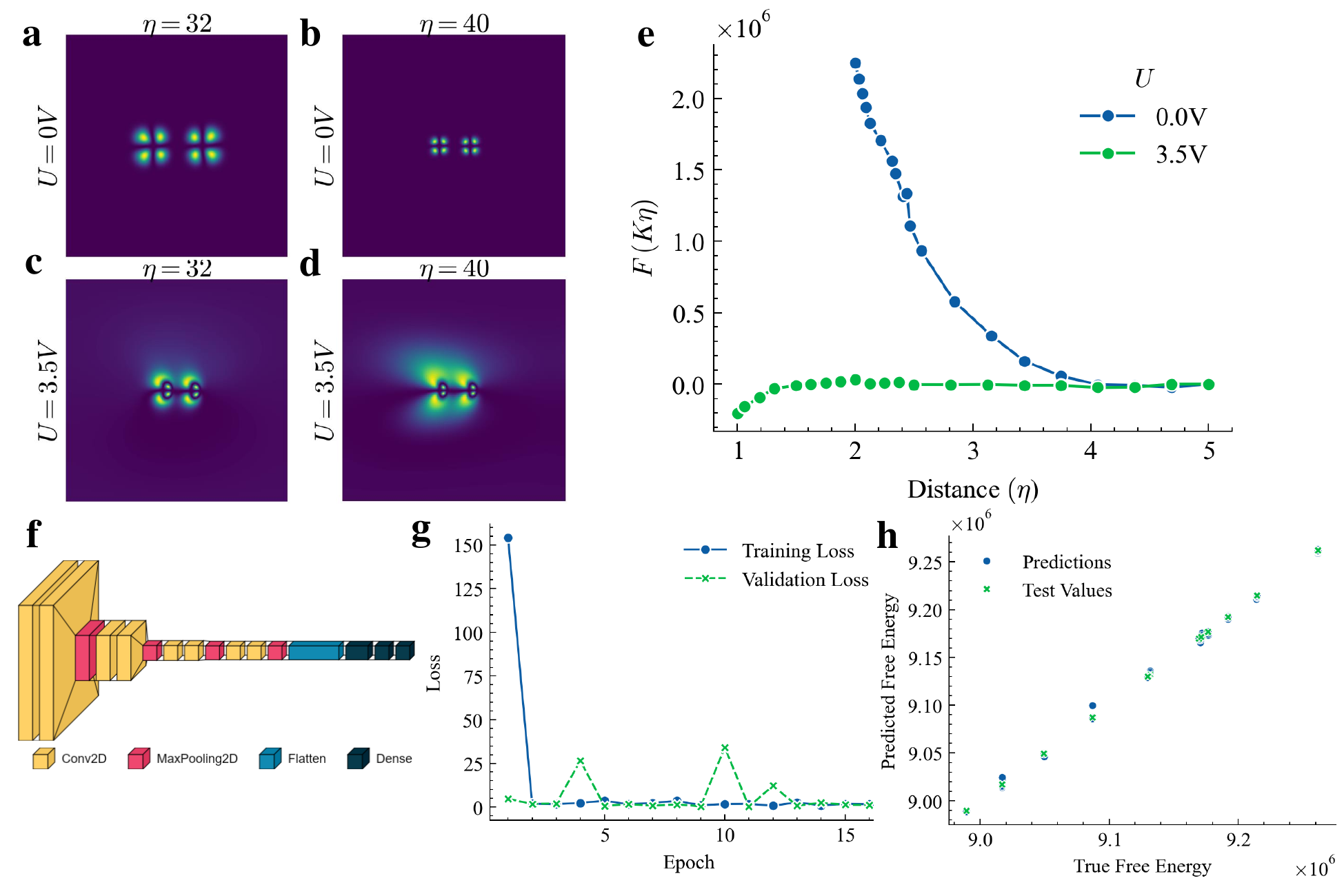}
    \caption{\textbf{a-d} POM images corresponding to the director configurations obtained by numerical minimisation of the Frank-Oseen free energy for two skyrmions at $\eta=32$ (\textbf {a, c}) and $\eta=40$ (\textbf {b, d}), under the presence of voltage $U=0V$ (\textbf {a, b}) and  $U =3.5 V$ (\textbf {c, d}) electric fields. \textbf{e} Frank-Oseen free energy as a function of the distance between two skyrmions at zero voltage (blue solid circles) and in electric field corresponding to $U=3.5V$ (green solid circles). \textbf{f} Schematic representation of the CNN architecture used for the classification of the POM images, and comprised of four blocks of two 4x4 convolutional layers with a 3x3 max-pooling layer, ReLU activation functions after all convolution operations, and a linear activation function in the output layer. \textbf{g} Training and validation losses as functions of training epochs. \textbf{h} Relationship between predicted and true energy values obtained by applying the training network to the test set (the crosses is the 1:1 relationship). The blue shaded area represents the standard deviation for each point.}
    \label{fig:3}
\end{figure*}

Figures~\ref{fig:3}\textbf{a-d} display the POM textures of two skyrmions at fixed separation, while varying both $\eta$ and $U$. These images reveal a complex dependence of the textures on both intrinsic material properties and external influences. Figure~\ref{fig:3}\textbf{e} plots the free energy as a function of the relative distance between two skyrmions. Notably, the interaction behaviour undergoes a shift from repulsive at zero voltage to attractive at $U=3.5V$. This shows significant role of external fields in modulating the skyrmion interactions.

\subsection*{Learning effective interactions between two skyrmions}

To predict the free energy of a system containing two interacting skyrmions, we have use the same CNN architecture as for the case of an isolated skyrmion free energy learning (see Fig.~\ref{fig:2}\textbf{e}). The efficacy of our approach is evident in the training and testing performance. Figure~\ref{fig:3}\textbf{f} shows that the loss function converges rapidly to a low value for both the training and test data, suggesting efficient learning and generalisation capabilities of the CNN. Furthermore, Fig.\ref{fig:3}\textbf{g} demonstrates the remarkable accuracy of our method, where the predicted free energy values exhibit a near-perfect 1:1 relationship with the true values. This exceptional agreement, reflected in a coefficient of determination of approximately 0.999, highlights the power of CNNs in learning to accurately predict the free energy of complex multi-skyrmion systems from their POM textures.

\subsection*{Machine learning from an input combining one- and two-skyrmion data}

\begin{figure*}[thb]
    \centering
    \includegraphics[width=0.9\textwidth]{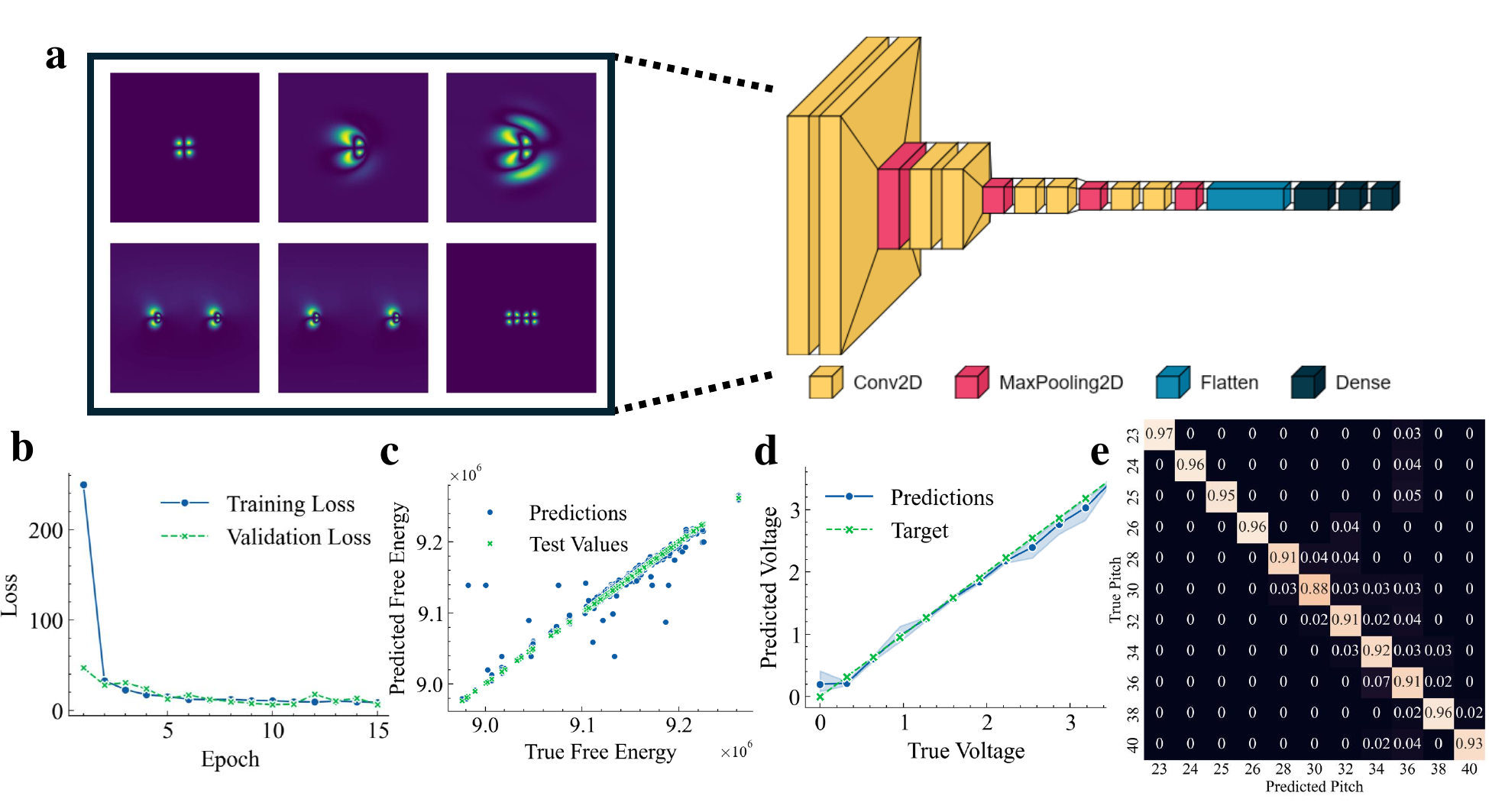}
    \caption{\textbf{a} Schematic illustration of the concept and showcasing of the simultaneous usage of one and two-skyrmion data to train the CNN. The CNN architecture is dependent on the output parameter as before. \textbf{b} Training (solid blue circles) and validation (green crosses) losses as a function of the number of epochs used during the training stage. 10\% of data has been separated as a test set, and the remaining is divided into training (85\%) and validation (5\%) sets. \textbf{c} Relationship between the predicted and the true free energy values obtained by applying the trained network to the test set (the crosses correspond to the 1:1 relationship). \textbf{d} Predicted voltage as a function of the true one. \textbf{e} Confusion matrix for the classification of the liquid crystal pitch $\eta$.}
    \label{fig:4}
\end{figure*}

A possibility to combine data from both single and multiple skyrmion configurations is not only a demonstration of the generalisability and versatility of our method but also a necessity for its practical application.  In realistic scenarios, LC devices may contain a varying number of skyrmions, and our model must be capable of predicting accurately properties across different configurations. Figure~\ref{fig:4}\textbf{a} illustrates this concept, showcasing the use of data from both one- and two-skyrmion configurations for the CNN training. This diverse dataset encompasses a range of LC properties, including the pitch, and external effects like the applied voltage, ensuring the model's versatility and robustness.

We demonstrate the effectiveness of this approach by predicting the free energy of different skyrmion configurations, regardless of whether they involve single or multiple skyrmions. Figure~\ref{fig:4}\textbf{b} depicts the loss functions depending on the training epochs, revealing a fast and clear convergence to a low value. This convergence indicates successful learning and generalisation across distinct skyrmion configurations. In Fig.~\ref{fig:4}\textbf{c}, we present the predicted free energy values against the true values, demonstrating a very good agreement. This finding emphasises the capability of our method to accurately estimate the free energy in diverse skyrmion systems.

The versatility of our approach extends beyond the free energy prediction. Figures~\ref{fig:4}\textbf{d} and \textbf{e} highlight the model's ability to predict both external perturbations and liquid crystal properties using data from multiple skyrmion configurations. Figure~\ref{fig:4}\textbf{d} shows the predicted applied voltage as a function of the true one, exhibiting a strong 1:1 agreement. This accuracy demonstrates the model's effectiveness in capturing the influence of external stimuli on skyrmion textures.  Furthermore, the confusion matrix for the classification of the LC pitch in Fig.~\ref{fig:4}\textbf{e} reveals a very good agreement as well, further confirming the model's ability to extract intrinsic material properties from complex POM images. These results  demonstrate collectively the potential of our approach for characterising and optimising LC systems with varying skyrmion configurations.

\section*{Discussion}

In this work we have shown the potential of Convolutional Neural Networks in predicting liquid crystal properties, such as the pitch and free energy, by analysing the optical textures of the skyrmions. We have shown that CNNs can accurately extract information concerning both the intrinsic material features and external stimuli from polarised optical microscopy images. Presented findings highlight the versatility of the developed approach. We have demonstrated that CNNs can accurately predict the pitch of a chiral nematic LC, the free energy of both single and multiple skyrmion configurations, as well as the applied voltage. The high accuracy achieved in these tasks, evidenced by the coefficients of determination exceeding 0.94 in all the cases, underscores the power of CNNs in deciphering complex patterns within POM textures and correlating them with relevant physical parameters.

This research opens up avenues for future applications, since the ability to rapidly and accurately characterise LC properties using CNNs can accelerate the development of new materials and devices. In particular, the precise control and understanding of material characteristics could lead to significant advancements in LC display technologies. Furthermore, the ability to predict the free energy from POM images can facilitate the study of skyrmion interactions and guide the design of novel skyrmion-based applications.

\section*{Methods}

\subsection*{Minimizing Frank-Oseen Free Energy}

We use the Frank-Oseen theory of liquid crystals where we write $F$ as the free energy for chiral nematic LC in the following form:
\begin{eqnarray}
    F=\frac{1}{2} \int_V \biggl [ K_1(\nabla\cdot\nvec)^2 + K_2 \left( \nvec\cdot\nabla\times\nvec-\frac{2\pi}{P} \right )^2 + \nonumber \\ K_3(\nvec\times\nabla\times\nvec)^2  + \varepsilon_0\Delta\varepsilon (\Evec \cdot \nvec)^2 \biggr ] d^3r,
    \label{eq:frank_oseen}
\end{eqnarray}
where $\nvec(\rvec)$ is the nematic director field, $K_1, K_{2}$ and $K_3$ are positive elastic constants describing splay, twist and bend director distortions, respectively, and $P$ is the cholesteric pitch, and the integral is taken over the three-dimensional (3D) domain $V$ occupied by the liquid crystal. We assume $V=L\times L\times L_z$, where $L_z$ is the separation between the confining rigid surfaces of the area $L \times L$. External electric field $\Evec$ couples to the nematic director according to the last term under the integral, where $\varepsilon_0$ is the vacuum permittivity and $\Delta\varepsilon$ is the dielectric anisotropy. The last term in the integrand of Eq.~(\ref{eq:frank_oseen}) approximates the local electric field by the constant external field, which is valid for weak fields, as used in the experiments \cite{Ackerman2017,Sohn2020}. In this study we assume that the electric field is applied in the direction  $\hat{\bm z}$ perpendicular to the confining cell walls and has a form $\Evec = -U \hat{\bm z}$, where $U$ is constant voltage drop across the cell.

The free energy \eqref{eq:frank_oseen} is minimised by using the following relaxation dynamics of the director field:
\begin{equation}
\partial_t n_\mu = -\frac{1}{\gamma} \frac{\delta F}{\delta n_\mu}.
 \label{director-time-eq}
\end{equation}
where $\gamma$ is the rotational viscosity, and $\mu=x,y,z$. Minimisers of \eqref{eq:frank_oseen} corresponds to steady state solutions of \eqref{director-time-eq}, which we compute numerically on a square lattice. The spatial derivatives on the r.h.s of Eq.~(\ref{director-time-eq}) are approximated by using finite-differences and the integration over time is performed using the fourth-order Runge-Kutta method. The director field at the confining surfaces $z = 0$ and $z = L_z$ is kept fixed perpendicular to the surfaces, and periodic boundary conditions are applied in the $x$ and $y$ directions. Equation (\ref{director-time-eq}) is integrated subject to the constraint $(\nvec \cdot \nvec) = 1$.  The values of the model parameters used in this study are provided in table \ref{table}.

\begin{table}[h]
\centering
\caption{\label{tab1} Parameters used in numerical integration of Eq.~(\ref{director-time-eq}) and physical units.}
\footnotesize
\begin{tabular}{@{}|c|l|l|l|}
\hline
symbol&sim. units & physical units&description\\
\hline
$\Delta x$&1 & 0.3125 $\mu$m& lattice spacing\\
$\Delta t$&1 & 10$^{-6}$ s& time step\\
$L$&300 & $ 94 \mu m$ & simulation box side length\\
$L_z$&32 & $10 \mu m$ & confining surfaces separation\\
$K_{1}$&17.2 &17.2$\times 10^{-12}$ N & splay elastic constant\\
$K_{2}$&7.51 &7.51$\times 10^{-12}$ N & twist elastic constant\\
$K_{3}$&17.2 &17.2$\times 10^{-12}$ N & bend elastic constant\\
$\gamma$&162&0.162 Pa s&director rotational viscosity\\
$\Delta\varepsilon$ & -3.7 & -3.7& dielectric anisotropy\\
\hline
\end{tabular}\\
\label{table}
\end{table}
\normalsize

To facilitate the formation of skyrmionic configurations we adopt as an initial condition the following axially symmetric \textit{Ansatz} for the director field 
\begin{align}
 &n_x(\rvec) = \sin\left (a(\rvec) \right ) \sin\left (b(\rvec) + \frac{\pi}{2} \right )\nonumber\\
 &n_y(\rvec) = \sin \left (a(\rvec) \right )  \cos \left (b(\rvec) + \frac{\pi}{2} \right )\nonumber\\
 &n_z(\rvec) = -\cos\left (a(\rvec) \right ) ,
 \label{toron_ansatz}
\end{align}
where (for $z' \equiv z-C_z $)
\begin{equation}
 a (\rvec)=  
 \begin{cases}
\frac{\pi}{2}\left[1 - \tanh\left(\frac{B}{2}(r-\sqrt{R^2-(z')^2})\right)\right],R>|z'|\\
0, R\leq|z'|
\end{cases}
\end{equation}
\begin{eqnarray}
 && b (\rvec)= \tan^{-1}\left(\frac{x-C_x}{y-C_y}\right)\\
 && r=\sqrt{(x-C_x)^2+(y-C_y)^2} .
\end{eqnarray}
 $(C_x,C_y,C_z)^{\mathrm{T}}$ above is the skyrmion center, $R$ set the skyrmion extention, $B$ controls the width of the twist wall of the skyrmionic tube, $r$ is the distance from a given point $(x,y,z)^{\mathrm{T}}$ to the skyrmion symmetry axis, and $b(\rvec)$ is the $2D$ polar angle. The values of the parameters used in the simulation are (in the simulation units defined in Table \ref{table}): $R=0.45L_z$, $B=0.5$, $C_x=C_y=L/2$, $C_z=Lz/2$.

\subsection*{Convolutional Neural Networks}

We have implemented a CNN approach, which is a special type of neural network designed for processing data with relevant spatial patterns \cite{Goodfellow2016}. Unlike traditional methods whihc require manual feature extraction, such as textures and shapes, a CNN directly processes raw pixel data from the image to extract pertinent features \cite{Simonyan2015}. 

At the core of a CNN is the convolution operation. The network takes an input feature map, represented as a three-dimensional matrix, and extracts tiles from it. Filters are applied to these tiles to compute new features, resulting in an output feature map (convolved feature) that may differ in size and depth from the input. These filters move across the input feature map grid both horizontally and vertically, one pixel at a time, extracting the corresponding tiles. Each convolutional layer performs this operation and passes the result to the next layer. The filter, or kernel, can be adjusted by modifying hyperparameters such as its size or stride. This process is analogous to the response of a neuron in the visual cortex, where each convolutional neuron processes information only from its receptive field. While fully connected feedforward neural networks could theoretically be used to extract features and classify data of this nature, they become impractical for large inputs like high-resolution images due to the massive number of neurons required, even in a shallow architecture.

Following each convolution operation, the CNN applies an activation function to the convolved feature, introducing non-linearity into the model. This non-linearity allows the network to adapt to various data types and to differentiate between different outcomes effectively.

A pooling layer is typically applied after the convolution operation to downsample the convolved feature map, reducing its dimensionality while preserving the most relevant feature information. This layer operates similarly to the convolutional layer by sliding over the feature map, extracting tiles, and performing a simple calculation on each, such as the maximum (max pooling) or average value (average pooling). Pooling helps summarize the presence of features, making representations invariant to small translations in the input data and improving computational efficiency. Pooling layers are often inserted periodically between successive convolutional layers in a CNN architecture. The rationale behind pooling is that the precise location of a feature is less important than its general location relative to other features.

The fully connected (FC) layer functions similarly to those found in conventional neural network models, operating on a flattened input where each input is connected to all neurons. The output from the convolution and pooling layers is flattened and fed into the fully connected layer, where numerous dot product operations are performed to produce the final output. FC layers are typically found toward the end of a CNN architecture and are used to optimize objectives such as class scores.

There are numerous ways to design a convolutional neural network, involving choices related to depth, width, number and size of filters, stride, and more. These decisions are often empirical, based on the input data type and inspired by successful architectures for similar tasks. Nonetheless, some design patterns are commonly found across CNN architectures \cite{Sigaki2020}. These include simplicity, symmetry, incremental feature construction, normalisation, and downsampling as the network depth increases \cite{Szegedy2015}. Our design choices, though guided by these principles, were largely determined through trial-and-error and cross-validation across several network parameters.

\bibliography{library}

\section*{Acknowledgements}

We acknowledge financial support from the Portuguese Foundation for Science and Technology (FCT) under Contracts no. PTDC/FIS-MAC/5689/2020 (https://doi.org/10.54499/PTDC/FIS-MAC/5689/2020), EXPL/FIS-MAC/0406/2021, CEECIND/00586/2017, UIDB/00618/2020 UIDB/00618/2020 (DOI 10.54499/UIDB/00618/2020),  and UIDP/00618/2020 UIDP/00618/2020 (DOI 10.54499/UIDP/00618/2020).

\end{document}